\documentclass[sigconf]{acmart}
\usepackage[utf8]{inputenc}
\usepackage{amsmath, amsthm}
\usepackage{booktabs} 
\usepackage{subcaption}
\usepackage{multirow}
\usepackage{verbatim}
\usepackage{comp_codes/tikzit}
\tikzstyle{green sq}=[fill={rgb,255: red,141; green,211; blue,199}, draw=black, shape=rectangle]
\tikzstyle{yellow sq}=[fill={rgb,255: red,255; green,255; blue,179}, draw=black, shape=rectangle]

\newcommand{\vvv}{\mathbf{v}}

\setcopyright{rightsretained}

\begin{document}

\copyrightyear{2021}
\acmYear{2021}
\acmConference[ICAIL'21]{Eighteenth International Conference for Artificial Intelligence and Law}{June 21--25, 2021}{São Paulo, Brazil}
\acmBooktitle{Eighteenth International Conference for Artificial Intelligence and Law (ICAIL'21), June 21--25, 2021, São Paulo, Brazil}\acmDOI{10.1145/3462757.3466066}
\acmISBN{978-1-4503-8526-8/21/06}

\ccsdesc[500]{Applied computing~Law}
\ccsdesc[300]{Applied computing~Document analysis}
\ccsdesc[300]{Information systems~Data mining}
\ccsdesc[500]{Information systems~Recommender systems}
\ccsdesc[300]{Computing methodologies~Natural language processing}

\keywords{citation recommendation, citation normalization, legal text, legal opinion drafting, neural natural language processing}

\title{Context-Aware Legal Citation Recommendation using Deep Learning}

\author{Zihan Huang}
\authornote{Authors contributed equally to the paper.}
\author{Charles Low}
\authornotemark[1]
\author{Mengqiu Teng}
\authornotemark[1]
\author{Hongyi Zhang}
\authornotemark[1]
\affiliation{%
  \institution{Language Technologies Institute\\Carnegie Mellon University}
  }

\author{Daniel E. Ho}
\author{Mark S. Krass}
\affiliation{%
 \institution{Stanford University}
 }

\author{Matthias Grabmair}
\authornote{Corresponding author (matthias.grabmair@tum.de). Current affiliation at TUM; work largely conducted while employed at SINC as part of adjunct affiliation with Carnegie Mellon University, Language Technologies Institute.}
\affiliation{%
  \institution{Department of Informatics\\Technical University of Munich}
  }
\affiliation{%
  \institution{SINC GmbH}
  }

\date{\today}

\begin{abstract}
    Lawyers and judges spend a large amount of time researching the proper legal authority to cite while drafting decisions. In this paper, we develop a citation recommendation tool that can help improve efficiency in the process of opinion drafting. We train four types of machine learning models, including a citation-list based method (collaborative filtering) and three context-based methods (text similarity, BiLSTM and RoBERTa classifiers). Our experiments show that leveraging local textual context improves recommendation, and that deep neural models achieve decent performance. We show that non-deep text-based methods benefit from access to structured case metadata, but deep models only benefit from such access when predicting from context of insufficient length. We also find that, even after extensive training, RoBERTa does not outperform a recurrent neural model, despite its benefits of pretraining. Our behavior analysis of the RoBERTa model further shows that predictive performance is stable across time and citation classes.
\end{abstract}

\maketitle

\section{Introduction}

Government agencies adjudicate large volumes of cases, posing well-known challenges for the accuracy, consistency, and fairness of decisions \cite{ames2020due,  mashaw1985bureaucratic}. One of the prototypical mass adjudicatory agencies in the U.S.\ context is the Board of Veterans' Appeals (BVA), which makes decisions on over fifty thousand appeals for disabled veteran benefits annually. Due to these case volumes and constrained resources, the BVA suffers from both a large backlog of cases and large error rates in decisions. Roughly 15\% of (single-issue) cases are appealed and around 72\% of appealed cases are reversed or remanded by a higher court \cite{Ho2019MassAdjudication}. These challenges are typical for agencies like the Social Security Administration, the Office of Medicare Hearings and Appeals, and the immigration courts, which adjudicate far more cases than all federal courts combined.  Lawyers and judges are hence in great need of tools that can help them reduce the cost of legal research as they draft decisions to improve the quality and efficiency of the adjudication process. 

Advancing the application of machine learning to suggesting legal citations is essential to the broader effort to use AI to assist lawyers. Citations are a critical component of legal text in common-law countries. To show that a proposition is supported by law, writers cite to statutes passed by a legislature; to regulations written by agencies implementing statutes; and to cases applying legal authorities in a particular context. Such is the importance of citations to legal writing that the traditional method of selecting law students to edit law journals has been a gruelling test on the correct format of legal citations \cite{youmakemesic}. Achieving performance on more difficult tasks, like text generation and summarization, depends on a sophisticated treatment of citations. 

This paper reports on experiments evaluating a series of machine learning tools for recommending legal citations in judicial opinions. We show that deep learning models beat ordinary machine learning tools at recommending legal citations on a variety of metrics, which suggests that the neural models have a stronger capability to exploit semantics to understand which citation is the most appropriate. 

We also demonstrate the importance of context in predicting legal citations. For ordinary text-based machine learning models with limited capacity for detecting semantic meaning, structured contextual metadata improves virtually all predictions. For deep learning models, the utility of structured metadata emerges only in sufficiently difficult settings where there may be a weaker semantic link between the input and the target, and only for certain models. Still, this result shows the potential importance of context to citation predictions. Deep learning models that are able to better incorporate contextual cues from semantic inputs are likely to outperform methods without such capabilities.

Because the BVA corpus has never been made available to the research community, we are releasing the text for single-issue decisions, with legal citation tokenization, case metadata, and our source code upon publication at: \url{https://github.com/TUMLegalTech/bva-citation-prediction}. We believe many other advances can be built on this as a benchmark for natural language processing in law.

\section{Related Work}

\subsection{Citation Recommendation}

Citation recommendation is a well-studied problem in the domain of academic research paper recommendation, as researchers seek help to navigate vast literatures in their fields. Many of the approaches are transferable to the legal context. They can be broadly categorized into \textit{citation-list based} methods, which characterize a query document by the incomplete set of citations it contains and provide a \textit{global} recommendation of citations relevant to the entire document, and \textit{context-based} methods, which take a particular segment of text from the query document and provide a \textit{local} recommendation that is relevant to that specific context \cite{ma2020review}.

\subsubsection{Citation-List Based Methods}

In this setting, the researcher is drafting a paper and has an incomplete set of citations on hand, and seeks to find additional relevant papers. 

An early approach in \cite{collab-filtering} applies collaborative filtering to this task. There, citing papers are ``users'' and citations are ``items.'' Given a new user, the algorithm locates existing users with similar preferences to the new user, and recommends items popular among the existing users. Matrix factorization methods project the sparse, high-dimensional user-item adjacency matrix onto a low-dimensional latent space and compare similarity in this latent space. For example, \cite{forest-for-trees} uses Singular Value Decomposition to find the latent space, and finds performance gains over ordinary collaborative filtering. 

Graph-based approaches treat research papers as nodes and citations as edges (directed or undirected), and use graph-based measures of relevance to find relevant nodes to an input set corresponding to the researcher's incomplete set of citations. Examples include the Katz measure \cite{link-prediction2007}, PageRank \cite{pagerank} and SimRank \cite{SimRank2002}. \cite{Gori2006ResearchPR} applies a topic-sensitive version of the PageRank algorithm by up-weighting papers in the incomplete set. \cite{Strohman07recommendingcitations} finds the Katz measure of node proximity to be a significant feature.

The citation-list based approach has its drawbacks. It puts the burden of creating a partial list of citations on the user. Attorneys who are new to veterans' law would face the well-known ``cold-start'' problem, where they have difficulty generating enough citations as input to receive quality recommendations. Second, attorneys drafting an opinion may be more interested in local recommendations relevant to their current section of work rather than global recommendations that are generally relevant to the entire case. Third, citation-list based approaches do not exploit the rich information contained in the textual context of each citation. 

\subsubsection{Context-Based Methods}

In this setting, the researcher inputs a span of text (the \textit{query context}), which can be a particular sentence or paragraph, instead of a list of citations. The system recommends \textit{local} citations relevant to this query context. 

Traditional information retrieval approaches directly compare the words in the query context to the words in the title, abstract, or full text of each cited document, and apply scoring models such as Okapi BM25 \cite{OkapiBM25} or Indri \cite{Strohman05indri:a} to arrive at a similarity score that is used to rank documents. However, as \cite{Ritchie2009CitationCA} observes, the full text of cited documents is often noisy and may not contain words similar to those used to describe the document as a whole. This problem is especially pertinent in law. Legal decisions and statutes sometimes lack informative titles, and the key legal implications are often buried in a mountain of other factual or procedural details.

Intuitively, we expect the span of text preceding or surrounding a citation (its \textit{citation context}) to contain useful information pertaining to the content of the cited document and the reason for citation. This information can then be used for retrieval. \cite{Ritchie2009CitationCA, Bradshaw2003} demonstrate that indexing academic papers using words found in their citation contexts improves retrieval. He et al. \cite{context-aware} develop this idea further by representing each paper as a collection of citation contexts, and then using a non-parametric similarity measure between a query context and each paper for recommendation. Huang et al. \cite{Huang2015NeuralProbModel} use a neural network to learn word and document representations to perform similarity comparison in that space. More recently, in a work most similar to our approach, Ebesu and Fang \cite{Ebesu2017NeuralCitationNetwork} directly train an encoder-decoder network with attention for context-aware citation prediction and find that adding embeddings representing the citing and cited authors improves predictions.

\subsection{Legal Citation Prediction}

Because of the importance of citations to legal writing \cite{law-search}, prior work has explored machine-generated recommendations for legal authorities  relevant to a given legal question.

A number of commercial tools claim to assist users in legal research using citations. Zhang and Koppaka \cite{paulZhangSemantics} describe a feature in LexisNexis that allows users to traverse a semantics-based citation network in which relevance is determined by textual similarity between citation contexts. Other commercial offerings include ROSS Intelligence \cite{ross}, CaseText's CARA A.I. \cite{caraAI} and Parallel Search \cite{casetext}, as well as Quick Check by Thomson Reuters \cite{quickcheck}. The methodology of such offerings is largely proprietary.

Winkels et al. \cite{toward-legal-recommender} develop a prototype legal recommender system for Dutch immigration law, which allows legal professionals to search a corpus by clicking on articles of interest; the system returns cases with the highest between-ness centrality with the article. In \cite{law-search}, Dadgostari et al. consider the task of generating a bibliography for a \textit{citation-free legal text} by modelling the search process as a Markov Decision Process in which an agent iteratively selects relevant documents. At each step, the agent can choose whether to explore a new topic in the original paper or to select a relevant paper from the current topic of focus. An optimal policy is learned using Q-learning. They find this adaptive algorithm to outperform a simpler method, based on proximity to the original document, on the task of retrieving U.S. Supreme Court decisions.

Other works \cite{fowlerNetworkAnalysis,KoniarisLegalAnalysis} have tangentially analyzed properties of legal citation networks, exploring measures of authority and relevance of precedents, as well as macro characteristics of the network, such as degree distribution and shortest path lengths. Sadeghian et al. \cite{sadeghianSemanticEdge} develop a system to automatically identify citations in legal text, extract their context and predict the reason for the citation (e.g., legal basis, exception) based on a curated label set.

\section{Data}

\subsection{The BVA Corpus}
The BVA corpus we use contains the full text of over 1 million appeal decisions from 1999 to 2017. Accompanying each decision is a set of \textit{metadata} derived from the Veterans Appeals Control and Locator System (VACOLS), which includes fields such as the decision date, diagnostic codes indicating the veteran's injuries, the case outcome, and an indicator for whether the case was subsequently re-appealed. Each case also contains one or more `issue codes,' which are hand-coded by BVA attorneys and categorize the key legal or factual questions raised (e.g., ``entitlement to a burial benefit''). This paper focuses on a subset of 324,309 cases that raise a single issue and have complete metadata, although our methods can be generalized to the full corpus.

\begin{table}
\begin{center}
\begin{small}
\begin{tabular}{lcp{0.2\linewidth}p{0.2\linewidth}}
\toprule
& \# Values & Most Frequent Class (\# cases) & Least Frequent Class (\# cases)\\
\midrule
Year & 19 & 2009 (22,801) & 2017 (3,651)\\
\hline
Issue Area & 17 & Service Connection for Bodily Injury Claims (38,956) & Increased Rating for Nerve Damage (2,921)\\
\hline
VLJ & 289 & Anonymized (6,159) & Anonymized (6)\\
\bottomrule
\end{tabular}
\end{small}
\end{center}
\caption{Summary Statistics of Corpus Metadata Variables.}
\label{table:summary-stats}
\vspace{-0.2in}
\end{table}

\begin{figure}
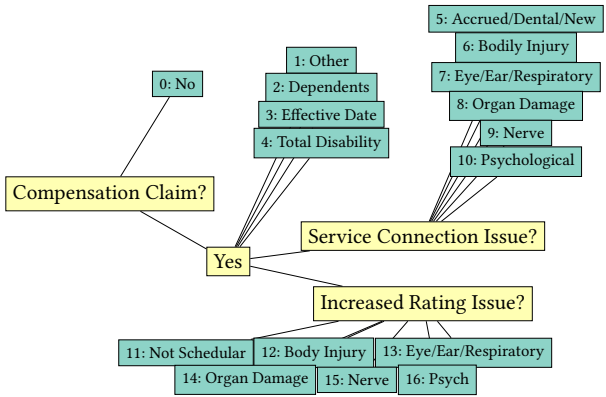

    \centering
    \ctikzfig{comp_codes/issuecodes}
    \vspace{-0.05in}
    \caption{Issue area categories. 
    }
    \label{fig:issarea}
    \vspace{-0.15in}
\end{figure}

We hypothesized that three metadata features would contribute to model performance. First, we included the \textit{year} of the decision, to reflect changes in citation patterns as new legal precedents emerge over time. Second, we constructed an \textit{issue area} feature to reflect the substantive issues presented in each case, which we hypothesize to provide strong priors for the type of citations contained within as well. The BVA has a hierarchical coding system comprising program codes, issue codes, and diagnostic codes to categorize each issue. For simplicity and class balancing, we curated a composite issue area variable with 17 classes (see Figure~\ref{fig:issarea}). Third, we included a feature referring to the Veterans' Law Judge (VLJ) who handled the case. This corresponds to the hypothesis that citation patterns vary with the idiosyncrasies of individual judges, inspired in part by \cite{Ebesu2017NeuralCitationNetwork}. Judge names were anonymized and judges with 5 cases or fewer were collapsed into a single unknown judge category. Summary statistics for these metadata are included in Table \ref{table:summary-stats}.

\subsection{Decision Text Preprocessing}
American legal citations follow a predictable format governed by \cite{bluebook}. Case citations, for instance, identify the parties to the case; the reporter containing the case; and finally the page in the reporter where the case begins. Thus, a citation to \textit{Brown v. Board of Education of Topeka} would begin as follows: \textit{Brown v. Board of Education}, 347 U.S. 483. This indicates that the first page of \textit{Brown} is found on the 483rd page of the 347th volume of the \textit{United States Reports}. The volume-reporter-page citation is usually a unique identifier for each case.\footnote{Summary dispositions of a case are sometimes reported in a table, such that multiple cases appear on a single physical page.} Citations to statutory law follow a similar three-part pattern: ``18 U.S.C. \S~46,'' means the 46th section of the 18th title of the United States Code. 

These three-part citation patterns form the basis for our text preprocessing pipeline. We first use a series of regular expressions to identify, clean, and classify citations from opinions. We then build a vocabulary of legal authority using publicly-available lists of valid cases and statutes. We use this vocabulary to extract all citations from case texts  and represent them using standardized indices. We describe this process in greater detail below.

\subsection{Citation Preprocessing}
\label{sec:citation-preprocessing}
The large raw citation vocabulary obtained from running regular expression extractors on every case is normalized into classes of case, statute, regulation, and unknown citations. 

For cases, this normalization involves matching the volume, reporter, and first/last page interval derived from the citation string with an authoritative list of cases found in the CaseLawAccess (CLA) metadata.\footnote{CLA is a public-access project that has digitized the contents of major case reporters \cite{caselawaccess}. We include the \textit{Vet. App.} and \textit{F.3d} reporters, which contain veterans' law cases and cases from the Federal Courts of Appeal, as these account for the vast majority of cases cited in the corpus.} If an extracted citation can be matched to a CLA metadata entry, it is replaced with a reference to that entry in the citation vocabulary during tokenization. For example, the extraction `Degmetich v. Brown, 8 Vet. App. 208 (1995)' is resolved to the normalized `Degmetich v. Brown, 8 Vet. App. 208, CLA\#6456776' (i.e. CLA metadata entry 6456776), which becomes an entry in the citation vocabulary that is used for all identifiable references to the same case.  Citations to the U.S. Code and to the Code of Federal Regulations are extracted using patterns based on the `<chapter> U.S.C. <tail>' and `<chapter> C.F.R. <tail>' anchors. The tail typically consists of one or more section elements, which we break into individual elements that each become their own normalized citation with the same anchor and chapter (e.g., `18 U.S.C. \S\S~46(a), 46(b) becomes the two entries `18 U.S.C. \S~46(a)' and `18 U.S.C. \S~46(b)'). All citations that cannot be normalized into either case, code, or regulation classes will form the `unknown' class. Once normalized, the vocabulary is further reduced by removing all citation entries which occur less than 20 times in the training cases and resolving them to an `unknown citation' token. This threshold was manually chosen as a suitable tradeoff between extensive coverage of citations and baseline frequency to enable the model to learn. 

The training data contains about 5M extracted citation instances comprising roughly 97k unique strings. Our normalization procedure reduces this to a citation vocabulary of size 4287, of which 4050 ($~\approx$ 94.5\%) are normalized (1286 cases, 870 statutes, 1894 regulations). The normalized entries cover about 98.5\% of citations occurring in the tokenized decisions. This reduction effect is primarily due to (a) complex statutory citations breaking apart into a smaller set of atoms, (b) different page-specific citations to a case getting collapsed into a single CLA entry, and (c) different forms of variation reduction (e.g., removal of trailing parentheses with years, etc.).

The final vocabulary is then used to normalize all citations encountered in case texts. Citation strings are extracted and replaced with a placeholder. The case/code/regulation procedure outlined above is applied to each citation string to obtain a list of one or more corresponding normalized citations. Each of these is kept if it is contained in the final vocabulary, or replaced with the `unknown citation' otherwise. The resulting sequence of vocabulary index tokens is re-inserted  at the location of the general citation placeholder after the text has been tokenized. Note that only citations containing reporter and page references are extracted and regularized. Short form citations (e.g., `id.') are treated as ordinary text and are excluded from the pool of prediction targets.\footnote{While this choice may limit the pool of prediction targets, it does not threaten the integrity of predictions themselves. By convention, short-form citations always follow full-form citations, which we detect. Because the system only has access to left context, it cannot `cheat' by reference to short-hand citations.} We also do not treat quotations in the text in any special way and rely on the tokenizer to capture them as part of the context window.

\section{Problem Definition}

We model the legal citation prediction problem as follows. Suppose a BVA staff attorney is drafting an opinion regarding an appeals proceeding. We refer to this document as $d$. The incomplete draft may already contain several citations to authority. We call this incomplete set $\mathbf{c}_d \subset C$, where $C$ represents the entire corpus of legal authorities, comprising possibly relevant cases, statutes, and regulations. The first task we consider is to predict the next citation $c^* \in C \setminus \mathbf{c}_d$ that is globally relevant to the opinion, given the incomplete set $\mathbf{c}_d$. This corresponds to the \textit{citation-list based} approach.
In our experiments, for a document that contains $M$ citations, we model the incomplete list $\mathbf{c}_d$ by taking the first $m$ citations $(1\leq m< M)$ from the document. We then seek to predict the next citation, i.e., the $(m+1)$-th citation $c^*$ given $\mathbf{c}_d$.

Alternatively, the attorney may be more interested in legal authority specific to the current segment of the opinion he/she is working on. We represent this text segment of interest by a sequence of tokens as the \textit{query context} $\mathbf{b}_d = \{b_1, ..., b_l\}$. The second task is thus to predict the next upcoming citation $c^* \in C$ that is locally relevant to context $\mathbf{b}_d$. This corresponds to the \textit{context-aware approach}. 
Specifically in our experiments, given a query context $\mathbf{b}_d$ of length $l$ in the document $d$, we seek to predict the first citation that occurs in the upcoming forecast window of length $w$. We vary length $l$ and forecast window $w$ depending on the method (see Section \ref{section:methods}).

In addition, metadata describing characteristics of the draft decision may also aid in citation prediction. For instance, the relevance and validity of case citations can change over time as new precedents emerge and others are overruled. Since many of the relevant legal standards are specific to particular classes of claims, the issue code feature may help identify relevant citations. Finally, different VLJs may have different propensities to cite certain authorities.

\section{Methods}
\label{section:methods}

Our main metrics are recall at 1, recall at 5 and recall at 20, that is, the proportion of data instances where the correct next citation is among the model's top 1, 5, and 20 predictions. Precision would not be an informative metric as we are only seeking to predict the single correct next citation. Recall at 1 reflects a restrictive user that expects the system to predict a single citation only. Recall at 5 simulates what we think is the typical user, who benefits from a small number of recommendations that can quickly be examined for the most appropriate one. A longer list of 20 simulates users seeking to get a bigger picture of what could possibly be relevant. We split the 324,309 single-issue BVA cases into 233,506 (72\%), 58,370 (18\%) and 32,433 (10\%) cases for the training, validation and test set, respectively. Each model is trained on the training set, tuned on the validation set, and tested on a 6-fold split of the test set to measure statistical uncertainty.

We implement four different methods on the task of legal citation prediction on the BVA corpus, and examine their comparative performance: a citation-only collaborative filtering system, a context-similarity based model, a BiLSTM recurrent neural classifier, and a RoBERTa-based classifier that has been pretrained on a language model objective. We note that our task is related to legal language generation (e.g., \cite{peric2020legal}). However, evaluating the citation prediction ability of a language generation model is significantly more difficult. Citations would need to be captured dynamically during a parameter-dependent generation process, validated, and resolved against the vocabulary. By contrast, the neural models in this project are implemented as conceptually straightforward classifiers, allowing us to test their ability to read the context well enough to forecast what will be cited next. We plan to tackle citation prediction as language generation in future work.

\subsection{Collaborative Filtering}
\label{section:collab-filter}

Our first experimental model uses \textit{collaborative filtering}, a common recommender system technique based on the assumption that similar users will like similar items. Transferred to our setting, each BVA decision document is treated as a user, and each citation is seen as an item. The prediction task then takes as input the citations that are already cited in a BVA draft opinion (which can be seen as the items that a user has liked), and returns other citations that similar documents have also cited.

Formally, assume that the corpus of BVA cases $C$ has $V$ authorities that can be cited. Then every document $d'$ can be represented by a sparse vector $\vvv_{d'}\in\mathbb{R}^V$, each of whose dimensions $\vvv_{d',c}$ indicates an importance score of a citation $c$ to the document. If citation $c$ is cited in a document, possible scoring functions could include a binary representation $(\vvv_{d',c}=1)$, a term frequency vector (tf), and a tf-idf vector that incorporates the inverse document frequency (idf). With such a representation, a set of document vectors $\{\vvv_{d'}:d'\in\mathcal{D}\}$ can be constructed from a document collection $\mathcal{D}$.

Given a draft of a BVA opinion $d$, its incomplete citation set $\mathbf{c}_d$ can also be summarized into a document vector $\vvv_{d}$. We use a collaborative filtering approach known as the user-based top-$K$ recommendation algorithm. The algorithm first identifies the $K$ documents $\mathcal{D}_K(d)$ that are most similar to $d$ from the collection based on their vector representations $\vvv$, based on their cosine similarity:
$$\mathrm{sim}(\vvv_{d},\vvv_{d'})=\frac{\vvv_{d}\cdot\vvv_{d'}}{\|\vvv_{d}\|_2\|\vvv_{d'}\|_2}.$$
The algorithm then finds candidate citations based on what these documents cite. An average of these document vectors weighted by their similarities gives the final recommendation. Specifically, the recommendation score of citation $c$ for document $d$ is given by
$$\mathrm{score}(d,c)=\frac{
\sum_{{d'} \in\mathcal{D}_K(d)} \mathrm{sim}(\vvv_{d},\vvv_{d'}) \vvv_{d',c}}{
\sum_{{d'} \in\mathcal{D}_K(d)} \mathrm{sim}(\vvv_{d},\vvv_{d'})
}.$$

In our experiments, the document vectors are collected from the training set. The number of top similar documents $K$ is a hyperparameter that can be tuned, and $K=50$ is chosen for the results reported. From our trials with three different scoring functions for the document vectors, binary scoring proved to be the most effective choice and was used throughout the experiments.

To incorporate metadata features, a score is assigned to each categorical feature $f_i$, namely the probability of citing the citation $c$ after conditioning on that feature:
$$
\mathrm{score}(f_i, c) = P(c\mid f_i).
$$

We take a weighted average of these features and the output of the collaborative filtering algorithm. We adopt the commonly used svmRank algorithm of \cite{clickthrough} to learn weights for each feature. We extract all citation occurrences in a random sample of 1000 documents from the training set, perform a pairwise transformation on the data, apply min-max normalization on the pairwise data, and train a linear Support Vector Machine (SVM) on the normalized data. The final score is a linear combination of individual feature scores using the learned weights. Citations suggested by the recommender system are reranked by their final scores and the top citations are chosen as final predictions.

\subsection{Text Similarity}

The second model uses a context-aware bag-of-words approach to predict citations. Previous studies, such as \cite{comparing-contexts, context-aware}, have demonstrated that the local context of words surrounding each citation occurrence can be used as a compact representation of the cited document to improve retrieval effectiveness, much like how in-link text is used to improve web retrieval. By contrast to collaborative filtering, this approach does not require the user to input an existing set of citations. Instead, the words in a section of interest within the draft opinion are used as a query to find the most relevant citation based on textual similarity of the present context to the previous contexts associated with each citation. Such local citation recommendations have the added advantage of relevance to a particular section of the opinion.

Formally, we adopt the approach of \cite{context-aware}, which represents each context by its tf-idf vector (normalized to have an L2-norm of 1). Each citation $c$ is represented by a collection of tf-idf vectors $\{\mathbf{b}_j: j=1, 2, \cdots, k_c\}$, where each $\mathbf{b}_j$ represents the local context of one citation occurrence and $k_c$ is the number of times $c$ was cited in the training set. Given a query context $\mathbf{b}_d$ at test time, the relevance of each citation $c$ to the query is then calculated as:
$$\mathrm{score}(\mathbf{b}_d, c) = \frac{1}{k_c} \sum_{j=1}^{k_c} 
        (\mathbf{b}_d \cdot \mathbf{b}_j)^2
$$

We removed stopwords, words that occurred in less than 10 documents, and words that contained digits. The most frequent 25,000 words were then chosen as a vocabulary. We used the 50 words preceding (instead of surrounding) each citation as its context, in line with our task to recommend relevant upcoming citations.\footnote{Note that this means citations are always the very next word after the context. This contrasts with the neural models presented below, where citations may appear at some distance from the context.} Citations that occurred within each context were also used as part of the vocabulary. As some citations were very frequently cited, we collected at most 100 randomly chosen context vectors (i.e. $k_c \leq 100$) per citation. Metadata features are incorporated into the model in a way similar to the Collaborative Filtering model (see Section \ref{section:collab-filter}). Each feature is assigned a score and an SVM model is trained to learn feature weights to produce the final score.

\subsection{Bi-directional Long Short Term Memory}
LSTMs \cite{hochreiter1997long} are a popular form of recurrent neural networks and serve as a well-known baseline for deep neural network models. Variants using LSTM remain competitive in various NLP tasks \cite{melis2017state, ma2018state, li2020attention}. BiLSTM (Bi-directional LSTM) improves on the original LSTM by reading inputs in both forward and backward directions. We adopted a two-layer BiLSTM on the BVA corpus for citation prediction. Just like the text similarity baseline, this approach performs local citation recommendation. It takes a sequence of words within the draft opinion as the query context, and predicts which citation is most likely to be cited next given the context. Going beyond the text similarity model, we predict the first citation that appears within a forecasting window of fixed length.

\begin{figure}[th]
\includegraphics[width=8.5cm]{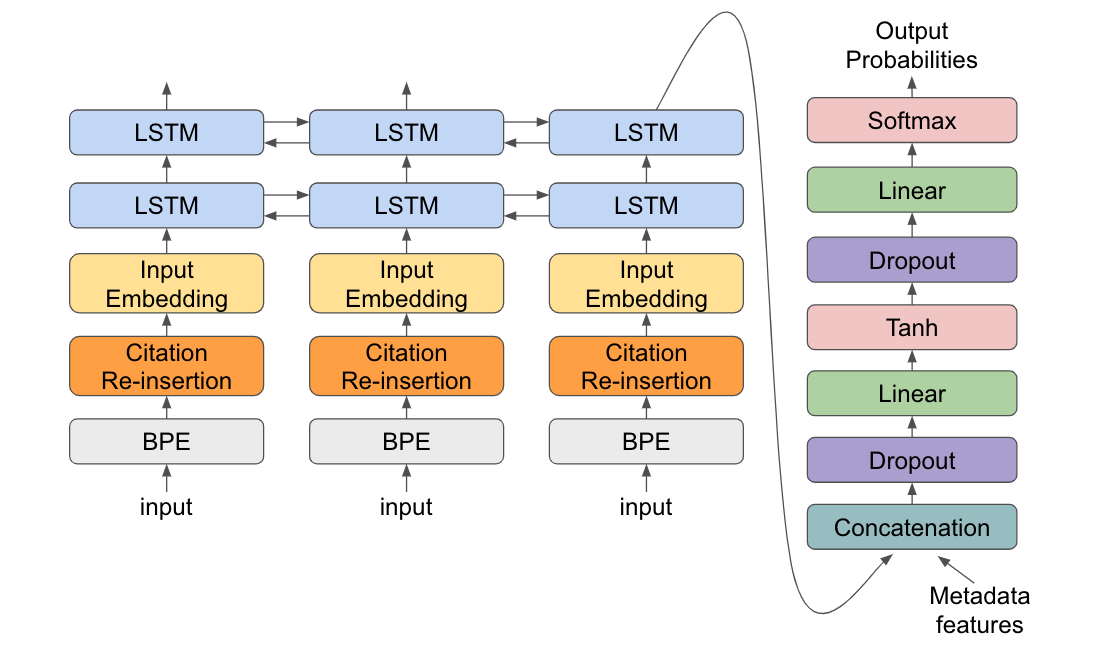}
\vspace{-0.25in}
\caption{The BiLSTM model architecture.}
\label{bilstm architecture}
\vspace{-0.13in}
\end{figure}

Formally, a sequence of tokens $\mathbf{b}_d = \{b_1, ..., b_l\}$ is extracted from each document $d$ as the query context and we seek to predict the immediate next citation in the upcoming forecasting window of length $w$. The query context is encoded using pre-trained byte-level Byte Pair Encoding (BPE) \cite{sennrich2015neural}. For comparability with the RoBERTa model, we use the `roberta-base' tokenizer provided by Huggingface \cite{wolf2019huggingface}, which has a vocabulary of about 50k tokens. The citation vocabulary indices are re-inserted after encoding, replacing the general citation token to generate the final encoded tokens as described in Section \ref{sec:citation-preprocessing}. The encoded tokens are fed into an embedding layer followed by two stacked bi-directional LSTM layers to produce a sequence of hidden states. The hidden state corresponding to the last token is used as the aggregate representation of the query context and flows into the classification head, which consists of two dropout/linear combination layers separated by a tanh activation, followed by a softmax layer to produce output probabilities for each citation, indicating how likely they will be cited next. Figure~\ref{bilstm architecture} illustrates the detailed architecture.

To incorporate the metadata information, processed metadata features are concatenated to the last hidden state before the classification layers as illustrated in Figure \ref{bilstm architecture}. The year and issue area features are one-hot encoded. The VLJ feature is projected into a three-dimensional vector space by feeding the VLJ ID into an embedding layer that can be inspected after training.

Our training setup follows common settings for language analysis experiments of this size. We use an embedding size of 768 and a hidden size of 3072. We compute CrossEntropy loss against a one-hot target vector of the same length as the vocabulary. To facilitate stable convergence, we use an effective batch size of 512 (implemented  via gradient accumulations across 4 batches of 128 to fit onto Nvidia P100 GPUs). We use an Adam optimizer with a fixed learning rate of $1e^{-4}$.

\subsection{Pretrained RoBERTa-based Classifier}
Since the introduction of BERT \cite{devlin2018bert}, language model pre-training has gained immense popularity, leading to models with superior performance on many NLP tasks and reductions in the amount of task-specific training data required. Its core mechanism is to compute a layer-wise self-attention across all tokens in the text, which allows it to effectively capture long-distance interactions without the architectural restrictions imposed by sequential models. RoBERTa \cite{liu2019roberta}  further improved BERT by employing certain techniques, such as longer training, and key hyperparameter adjustment. We apply this model to our task via transfer learning to test how a Transformer model pretrained on a language model objective performs against our BiLSTM model trained from scratch.

We fine-tuned a pre-trained RoBERTa model (HuggingFace's `roberta-base' \cite{wolf2019huggingface}) on the BVA corpus using the citation prediction task. The model uses 24 layers, a hidden size of 1024, 16 self-attention heads, leading to 355M parameters overall. We apply a common sequence classification architecture and, similar to our BiLSTM model, feed the final hidden layer's output through two dropout/linear layers separated by a tanh activation to produce the final hidden vector $C$ in the citation vocabulary size.

\begin{figure}[t]
\includegraphics[width=8.0cm]{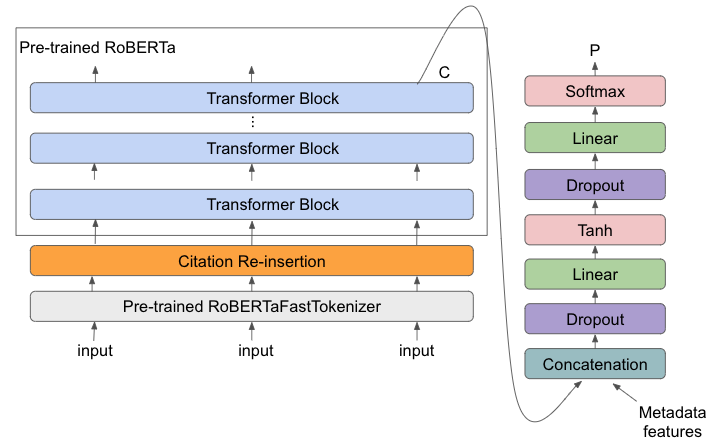}
\vspace{-0.25in}
\caption{The RoBERTa model architecture.}
\label{RoBERTa}
\vspace{-0.13in}
\end{figure}

To fine-tune our RoBERTA model, we use the same data preprocessing and loading as in the BiLSTM experiment. We tokenize the BVA decisions with the pre-trained RoBERTa tokenizer provided by Huggingface \cite{wolf2019huggingface} and apply our citation extraction and normalization procedure. Sequences are padded to the same length and an attention mask is generated to indicate whether the corresponding token is a padding token. Formally, a pair of tensors ($\mathbf{b}_d$, $\mathbf{a}_d$) is extracted from each document $d$, where $\mathbf{b}_d$ represents the token ids and  $\mathbf{a}_d$ represents attention mask. Label $\mathbf{l}_d$ is the index in the citation vocabulary of the first citation following the given context $\mathbf{b}_d$. We compute cross-entropy loss between predictions $\mathbf{p}_d$ for ($\mathbf{b}_d$, $\mathbf{a}_d$) and label $\mathbf{l}_d$.

Again, to allow training with relatively large batches on NVidia P100 GPUs, we use a batch size of 192 and accumulate gradients for three steps before performing back-propagation, resulting an effective batch size of 576. We use the AdamW optimizer with a learning rate of $1e^{-4}$.

\subsection{Sampling-based Data Loading}
\label{section:data-loading}
Each data instance for the BiLSTM and RoBERTa models consists of a context window and forecasting window to the left and right side of an offset. During every training epoch, and during evaluation, we sample a random offset for each case from all offsets whose forecasting window contains a citation token. We designed data loading this way to mitigate the prohibitively large space of traversing all possible context/forecasting window combinations for all citations in all cases. Note that, because the target is always the first citation within the forecasting window, our data loading is biased against citations that rarely appear first in strings of successive citations. We plan to address this imbalance in future work. 

\section{Results and Discussion}

\begin{table*}[t]
\begin{center}
\begin{small}
\begin{tabular}{clcccccc}
\toprule
Model & Setting & \multicolumn{2}{c}{Recall@1} & \multicolumn{2}{c}{Recall@5} & \multicolumn{2}{c}{Recall@20}  \\
\midrule
Majority Vote & Original & 1.73\% & (.02\%) & 7.35\% & (.03\%) & 26.4\% & (.02\%) \\
\midrule
\multirow{4}{*}{\shortstack{Collaborative\\Filtering}}
& Original & 10.2\% & (.06\%) & 25.5\% & (.07\%) & 45.4\% & (.08\%) \\
& Original + Year & 9.68\% & (.05\%) & 24.8\% & (.08\%) & 45.2\% & (.09\%) \\
& Original + Year + IssueArea & 9.64\% & (.05\%) & 24.7\% & (.08\%) & 45.2\% & (.08\%) \\
& Original + Year + IssueArea + VLJ & 9.60\% & (.05\%) & 24.7\% & (.09\%) & 45.2\% & (.09\%) \\
\midrule
\multirow{4}{*}{\shortstack{Text\\Similarity}}
& Original & 16.4\% & (.03\%) & 41.1\% & (.04\%) & 66.2\% & (.05\%) \\
& Original + Year & 20.4\% & (.05\%) & 48.2\% & (.05\%) & 79.5\% & (.03\%) \\
& Original + Year + Class & 16.2\% & (.06\%) & 51.6\% & (.07\%) & 82.6\% & (.05\%) \\
& Original + Year + Class + VLJ & 16.6\% & (.06\%) & 51.7\% & (.06\%) & 82.7\% & (.05\%) \\
\midrule
\multirow{2}{*}{\shortstack{BiLSTM}}
& no metadata (47 epochs) & 65.2\% & (.33\%) & 81.8\% & (.14\%) & 91.1\% & (.11\%) \\
& all metadata (50 epochs) & 65.8\% & (.35\%) & 82.4\% & (.26\%) & 91.3\% & (.16\%) \\
\midrule
\multirow{2}{*}{\shortstack{RoBERTa}}
& no metadata (106 epochs) & 65.6\% & (.33\%) & 82.8\% & (.31\%) & 91.7\%& (.21\%) \\
& all metadata (126 epochs) & 66.2\% & (.30\%) & 83.2\% & (.17\%) & 92.1\%& (.20\%) \\
\bottomrule
\end{tabular}
\end{small}
\end{center}
\caption{Prediction results. Each model is evaluated on six folds of the test set and the numbers reported are the mean and the standard error (in parentheses) of recall at 1, 5, and 20. Neural models are trained using 256/128 context/forecast windows. All metadata includes year, issue area, and VLJ identifiers.}
\label{table:full-citation}
\vspace{-0.15in}
\end{table*}

\subsection{General Performance}

Table~\ref{table:full-citation} shows the full citation prediction results of the four models. We add a naive majority vote baseline, which always recommends the 20 most popular citations in descending order of their number of occurrences in the training data.

We first turn to our ordinary machine learning models. A comparison to their `original' setting --- without access to structured metadata --- shows the importance of semantic context for citation prediction. The collaborative filtering model uses only the previous citations in a document as input. It returns the correct citation as its top-ranked recommendation 10.2\% of the time; recall@5 is 25.5\%. By contrast, the text similarity baseline achieves a recall@1 of 16.4\% and a recall@5 of 41.1\%, on average. This is strong evidence that the textual context preceding a citation is a critical signal. By contrast, the document-level statistical information on citation patterns leveraged by collaborative filtering is less informative.

For the text similarity model, adding metadata information generally gives a noticeable improvement over predictions based on text alone. For example, adding structured information on the year of a decision improves performance, which suggests that the model does not otherwise detect temporal information. But not all metadata is equally useful. Adding information on the identity of the judge produces little or no marginal gain. Further, we do not find evidence that metadata enhances the collaborative filtering model. Interestingly, the benefit in recall@1 of case year information is negated when class is added, although recall@5 and recall@20 improve at the same time. If one were to pursue the baseline further, this effect should be examined.

For purposes of this comparison experiment, we train our BiLSTM and RoBERTa models on a context window of 256 tokens and a forecast window of 128 tokens. They are trained until, in our assessment, validation metrics indicated convergence, at which point they dramatically outperform both baselines. Both predict the correct citation roughly 65-66\% of the time and produce a recall@5 of around 81-83\% using the textual context alone. The neural models' improvement over the text similarity baseline suggests that the ability to encode more complex semantic meanings---and track long-term dependencies across context windows of significant length---noticeably improves performance in citation recommendations.

We experimented with different metadata combinations for the neural models with 8 epochs of training time and observed no clear differences, and decided to only train all-meta and no-meta models until convergence. Giving the BiLSTM and RoBERTa models access to metadata improves predictive performance by around 0.2-0.6\%. That delta, however, is mostly within two standard errors of the two models. Our two possible explanations are (a) that the neural models are capable of implicitly inferring some background features from the legal text itself, and thus they will not benefit much from us providing this information explicitly, and (b) that metadata may not carry much signal for this task.

The superior neural model performance is intuitive in legal text also because the text preceding a citation will typically paraphrase a legal principle or statement that is reflective of that source. We can assume that some portion of our context-forecast instances consist of relatively easy examples. To some degree, short-distance citation prediction can in fact be considered a sentence similarity task. Commercial search engines even use text encoding similarity to suggest cases to cite for a particular sentence (e.g., \cite{casetext}). Similarly, literal quotations from the source preceding the citation can be certain indicators. However, a pure memorization approach will fail for longer forecast distances, as one can anticipate an upcoming cited source from the narrative progression in the text before it becomes lexically similar to the source closer to the citation. An exception to this consists of large spans of boilerplate text that contain citations and are reused across decisions. To investigate the capacity of our models to anticipate citations from further away, we experiment with different forecasting lengths (see Section \ref{section:window-sizes} and \ref{section:error-analysis} below).

A final observation is the stability of predictive performance across the six test set folds as evidenced by the low standard errors. The neural models have slightly more deviation than the baselines and the BiLSTM and RoBERTa models metric are generally within the reach of $\pm 2$ standard errors within a given recall metric.

\subsection{Context \& Forecasting Window Sizes}
\label{section:window-sizes}

\begin{figure*}[ht]
\includegraphics[scale=0.75]{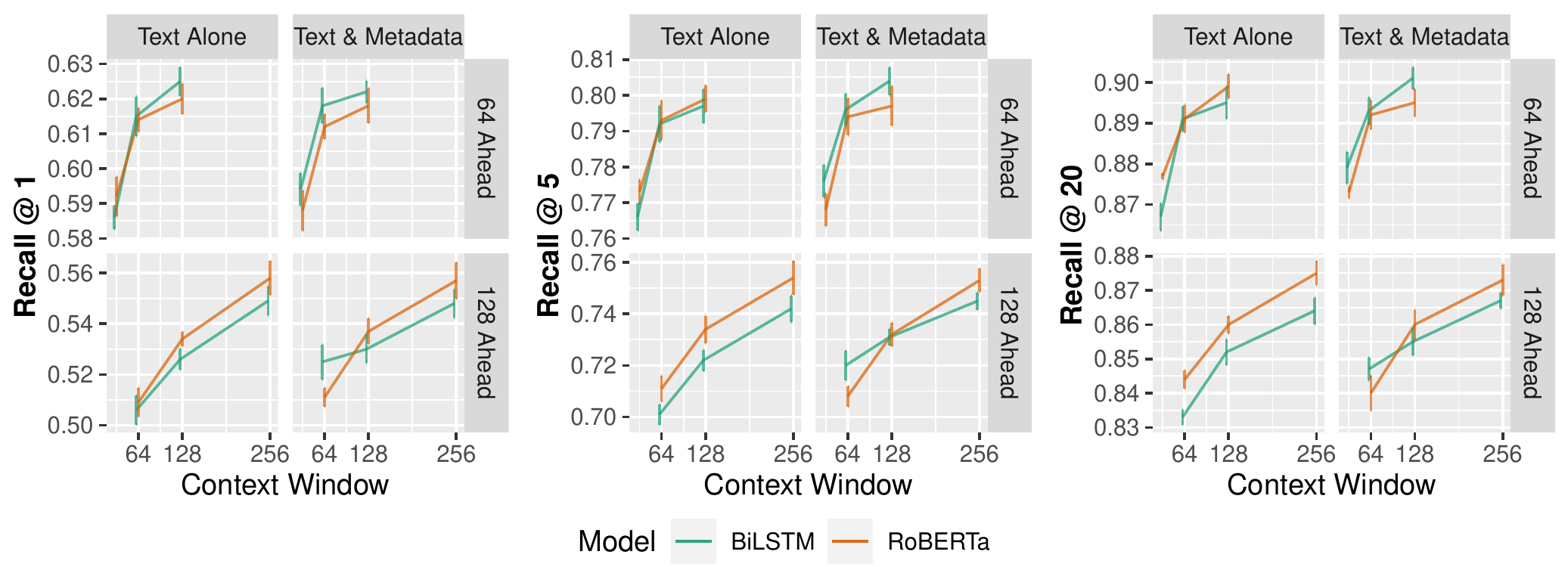}
\vspace{-0.15in}
\caption{Results of the ablation study for Recall at 1, 5, and 20. Within each panel, the most difficult tasks are in the bottom left corner and the easiest tasks are in the top right. The x-axis shows the context window. ``64 ahead'' and ``128 ahead'' refer to the maximum number of tokens between the context window and the target citation. Error bars are 95\% confidence intervals.}
\label{fig:ablation}
\vspace{-0.1in}
\end{figure*}

To further explore the behavior of the deep neural models, we conducted an ablation study, in which we varied the size of the context and forecasting windows and varied the availability of structured metadata information.
We tested 12 different settings for BiLSTM and RoBERTa, respectively. In this grid-search experiment, all models were trained for 8 epochs before test metrics were computed. The detailed settings and the results are illustrated in Figure \ref{fig:ablation}. 

As expected, increasing the forecasting window hurts performance by weakening the semantic link between input and target. Also unsurprising is the upward slope in each panel, which simply shows that providing the models with more context generally improves predictions. But the utility of added context changes with the difficulty of the forecasting task. When the target citation is nearer to the context (`64 ahead'), we observe diminishing returns to context: A 128-token context window is only slightly better than a 64-token context window. When the target is further away (`128 ahead'), more context helps. We hypothesize that adding additional context helps to compensate for the difficulty of the task: the network models are able to infer more clues for the citation given the extra information. As the size of the forecasting window increases, the potential for a weaker semantic relationship between the immediate context and the eventual citation makes it more helpful for the model to have access to additional context.

We observe a similar story with respect to structured metadata: the harder the task, the more helpful it is to add metadata. In the BiLSTM framework, metadata is most helpful when the model is given little context and when the targets are far away. But when the target is nearby (`64 ahead'), performance is statistically indistinguishable between models that do have access to metadata and those that do not. The findings align with our hypothesis that, when enough context is given, the neural network models are able to derive the clues for citations from text snippets, and thus obviating the need for metadata information.

\begin{figure}[ht]
\vspace{-0.15in}
\includegraphics[scale=0.5]{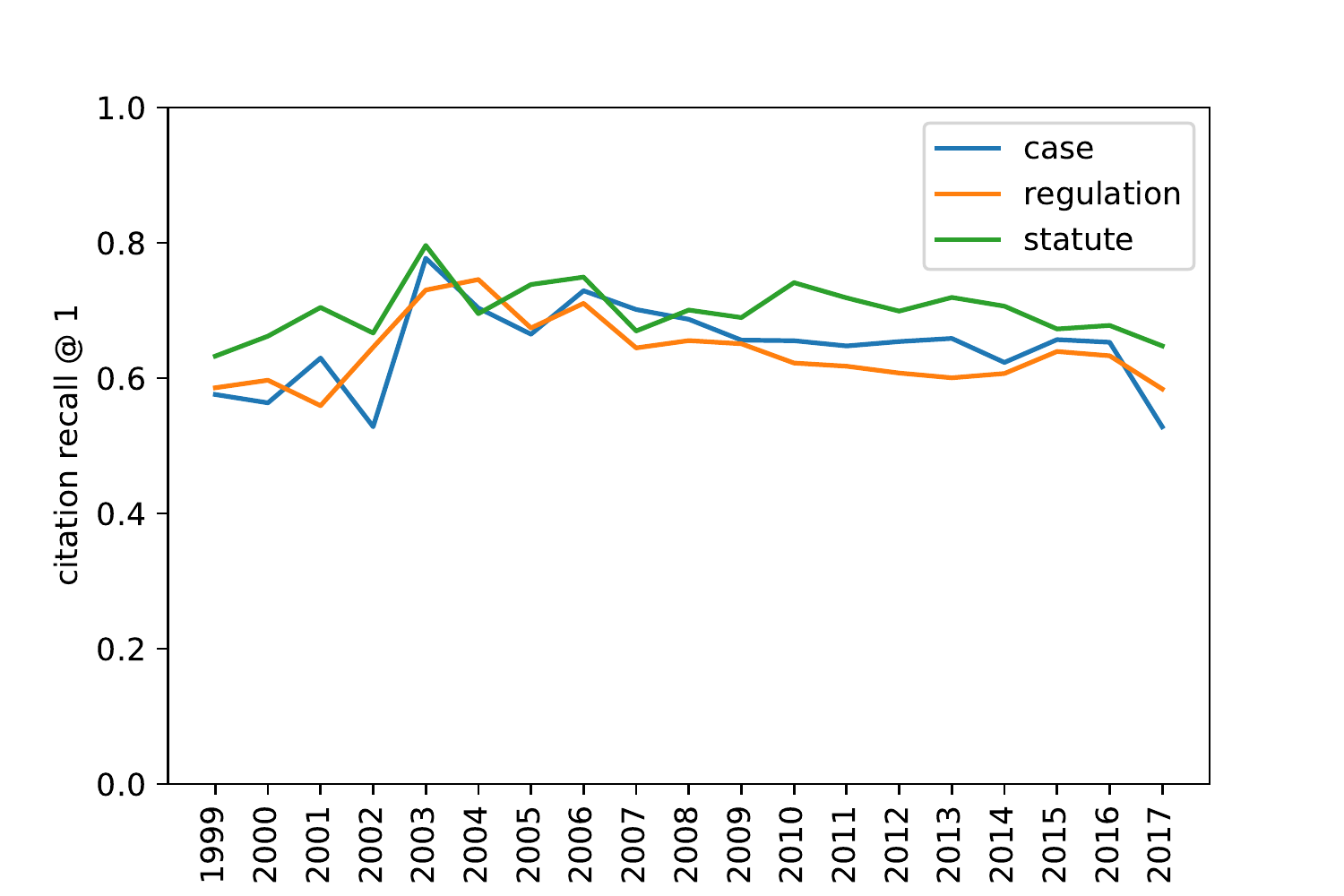}
\vspace{-0.1in}
\caption{Per-class recall@1 for RoBERTa all-meta model over time.}
\label{fig:per_class_recall_at_1}
\vspace{-0.1in}
\end{figure}

Although the ablation experiment was conducted with only 8 epochs of training, our experimental results on models trained until convergence are in line with the observation that the effect of metadata is only marginal. At the very least, however, our results indicate that a conclusive exploration of the effectiveness of individual metadata features in training neural citation recommenders may require considerable computational resources or the use of advanced techniques to reduce training time. Even after 126 and 50 epochs, respectively, our models showed no signs of overfit and the decision that the loss decrease had slowed down enough to stop training was a matter of judgment. Given the carbon footprint of neural model training \cite{strubell-etal-2019-energy}, we believe such ablation research on large neural models should be conducted with care.

\subsection{Pre-Training vs. Training From Scratch}

Despite its general English language-model pretraining, RoBERTa models do not show noticeable superiority over BiLSTM in the ablation experiments, even when a more challenging task and a more complicated context is given. For the models trained until convergence in Table~\ref{table:full-citation}, RoBERTa performs better than the BiLSTM model, but only by at most ~1\% and often with overlapping $\pm 2$ standard errors. One possible explanation is that the pretraining of RoBERTa is performed on non-legal text, negating the pretraining benefit for this domain-specific task. Alternatively, the task may not require sophisticated language understanding and/or our supervised setup provides sufficient training to learn citation prediction from scratch. We leave an exploration of the effects of domain-specific pretraining (e.g. using \cite{legalbert}) in this task for future work.

\begin{figure}[ht]
\vspace{-0.1in}
\includegraphics[scale=0.5]{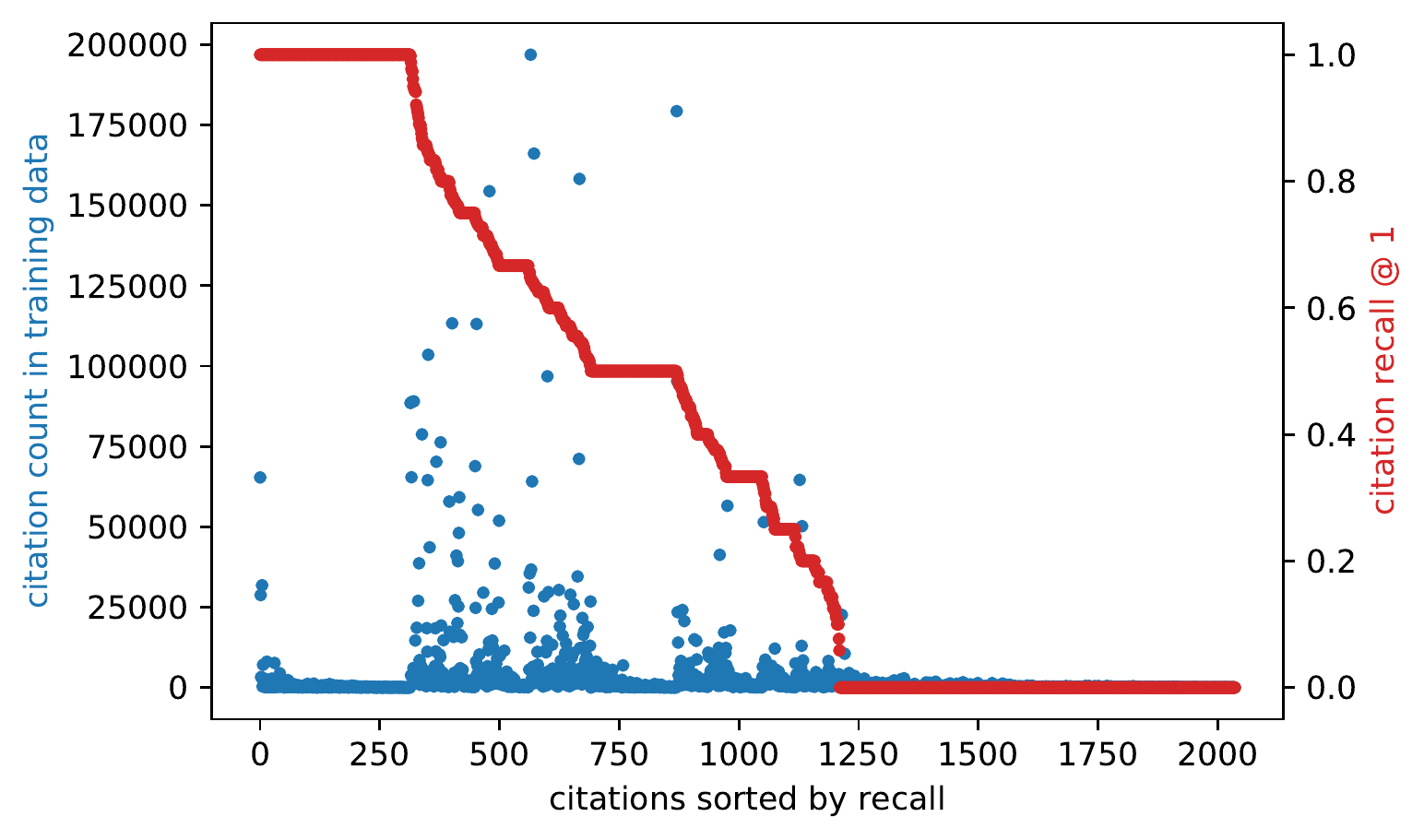}
\vspace{-0.15in}
\caption{Per-citation recall@1 vs. number of instances in training data for RoBERTa all-meta model.}
\label{fig:recall_at_1_vs_count}
\vspace{-0.15in}
\end{figure}

\subsection{Error Analysis}
\label{section:error-analysis}

Figure \ref{fig:per_class_recall_at_1} shows a relatively consistent recall at $k=1$ performance across classes over time. We see a slight downward slope for the case and regulation metric towards the end of our analysis period. This may be due to opinions later in the time period potentially containing new citations and patterns occurring less frequently in the training data. The plot exhibit a single strong upwards oscillation in 2002-2003. We believe this is likely due to litigation surrounding the Veterans Claims Assistance Act of 2000, which sparked mass remands by the BVA back to regional offices. This relative shape of the per-class recall graphs stays roughly the same for larger values of $k$, albeit shifted to higher absolute recall levels.

To assess the influence of the sampling distribution, the combined scatterplot in Figure \ref{fig:recall_at_1_vs_count} plots the recall at $k=1$ achieved for each citation against its frequency as a prediction target in the training data. Of the 2037 different citations that were loaded in a single pass over the test data (of the total of 4287; see Section \ref{section:data-loading}), only about 1200 citations are predicted with non-zero recall. At $k=20$ this number increases to about 1700 and the red curve shifts right (not shown). The distribution of blue data points indicates that almost all zero-recall citations occur with very low, or zero, frequency. However, citations with high recall do not follow a recognizable frequency pattern. This is informative for the cold-start problem of new sources becoming available that have not been cited enough yet to be learned by models such as the ones presented here. We are aware of this limitation and leave it for future work.

Finally, we examined whether the number of decisions in the test data authored by a judge correlated with the model's performance in predicting citations from those decisions, but did not find clear patterns. The three-dimensional judge embeddings also did not reveal any clear separation with regard to the per-judge recall. We intend to investigate the relationship between attributes of individual VLJs and the behavior of trained models in future work.

To help characterize the underlying behavior of the models, we drew a sample of 200 erroneous predictions generated by a long-trained RoBERTa model similar to the one in Table~\ref{table:full-citation}.\footnote{After qualitative error analysis was completed, a pre-processing bug was corrected, leading to changes in recall values of less than 0.5\%. Quantitative results and analyses of converged models reported here are from this slightly improved version.} Two sets of observations indicate that the model has developed some conceptual mapping of citations. First, 16\% of the erroneous predictions \emph{did} appear in the forecast window, somewhere after the first citation. Idiosyncrasies in citation order might explain these errors, but there is no conceptual mismatch. Second, somewhere around 5\% of the errors involve regulations that implement a particular statute. For example, one case cites 38 C.F.R. \S~3.156(a), a regulation defining when veterans may present ``new and material evidence'' to reopen a claim. The model predicted a citation to 38 U.S.C. \S~5108(a), which is precisely the statute commanding the BVA to reopen claims when veterans identify ``new and material evidence.'' Again, the erroneous prediction is in exactly the right conceptual neighborhood.

\begin{table}[t]
\begin{center}
\begin{small}
\begin{tabular}{cccccccc}
\toprule
Distance & $N$ & Recall@1 & Recall@5 & Recall@20 \\
\midrule
1-16 & 13609 & 78.7 & 91.9 & 97.1 \\
17-32 & 11237 & 75.6 & 89.8 & 95.9 \\
33-48 & 9125 & 68.8 & 85.2 & 93.2 \\
49-64 & 7082 & 63.3 & 81.1 & 91.0 \\
65-80 & 5452 & 55.6 & 75.4 & 87.6 \\
81-96 & 4534 & 52.7 & 73.1 & 87.1 \\
97-112 & 3918 & 47.9 & 69.6 & 84.0 \\ 
113-128 & 3403 & 42.1 & 66.2 & 82.7 \\
\bottomrule
\end{tabular}
\end{small}
\end{center}
\caption{Roberta all-meta performance binned by token distance from beginning of forecasting window to target citation, based on single pass over validation set.}
\label{table:forecast-analysis}
\vspace{-0.2in}
\end{table}

Consistent with our ablation analysis, our review of the errors suggests the critical role that topical changes in long texts play in generating errors. Table \ref{table:forecast-analysis} shows recall metrics for targets binned by the position of the target citation within the forecast window between minimum and maximum distances. Since legal analysis is often addressed in a single section of an opinion, close citations are more frequent than distant ones. Unsurprisingly, performance decreases with distance from the context window. From closest to farthest bin, recall@1 shrinks by a relative ~47\%, recall@5 by ~28\%, and recall@20 by ~15\%. This behavior is intuitive and indicates that the system may indeed memorize contexts immediately surrounding citations. Still, the gradual decline in performance, especially for recall@5, suggests that the model is learning some amount of longer-distance patterns. This forms evidence that effective citation recommendation benefits from both a sophisticated representation of context and supervised training on existing citation patterns.

\section{Conclusion}

In this paper, we have implemented and evaluated four models that can recommend citations to lawyers drafting legal opinions. BiLSTM and pretrained RoBERTa perform comparably and outperform the collaborative filtering and bag-of-words baselines. Our ablation experiments show that (a) adding metadata about case year, issue, and judge only leads to insignificant performance improvements for the neural models, and (b) predicting citations further away from the context is more difficult, which can be compensated to some degree by providing more context. Training for extended periods continuously improves up to a recall@5 of 83.2\%. As such, we have shown that context-based citation recommendation systems can be implemented as classifiers for a largely normalized citation vocabulary with acceptable performance. Further, our error analysis shows that even incorrect predictions may still be useful. 

Our work also points to the next steps for legal citation prediction. First, citation prediction can be conceived of more broadly as language generation. Research should hence explore whether neural models can go beyond pointing to an entry in the citation vocabulary and write valid citation strings appropriate for a given context, possibly as part of a continuation of the text. Second, as a practical matter, it will be important to evaluate the usefulness of the models trained here with expert users. Finally, we note that legal sources and institutions form dynamic systems. Constant adaptation, such as detecting and accounting for changes in precedent, will be key to the future utility of citation systems.

These future directions could rapidly improve legal citation, and our results here show that context-aware citation prediction can play a significant role in improving the accuracy, consistency, and speed of mass adjudication.


\section{Statement of Contributions}
The project was conceived and planned by all authors. ZH, CL, MT, and HZ conducted all model development and experimental work under the mentorship of DEH, MK, and MG. MK and MG developed the citation preprocessing functionality, as well as produced the error analysis. All authors contributed to writing the paper.

\section{Acknowledgments}
The authors thank CMU MCDS students Dahua Gan, Jiayuan Xu, and Lucen Zhao for creating the issue typology, Anne McDonough for supporting contributions around citation normalization, and Dave Ames, Eric Nyberg, Mansheej Paul,  and RegLab  meeting participants for helpful feedback.

\bibliographystyle{ACM-Reference-Format}
\bibliography{bibliography}

\end{document}